\documentclass[twocolumn,showpacs,amsmath,amssymb]{revtex4}
\usepackage{graphicx}
\usepackage{dcolumn}
\usepackage{bm}
\usepackage[dvips]{epsfig}

\newcommand{\ba}{\begin{eqnarray}}
\newcommand{\ea}{\end{eqnarray}}
\newcommand{\ban}{\begin{eqnarray*}}
\newcommand{\ean}{\end{eqnarray*}}
\newcommand{\be}{\begin{equation}}
\newcommand{\ee}{\end{equation}}
\newcommand{\bd}{\begin{displaymath}}
\newcommand{\ed}{\end{displaymath}}

\newcommand{\n}[1]{\label{#1}}

\newcommand{\non}{\nonumber}
\newcommand{\eq}[1]{(\ref{#1})}

\newcommand{\pa}{\partial}

\newcommand{\hhh}{\, ,\hspace{0.2cm}}

\begin{document}



\title{`Hidden' Symmetries of Higher Dimensional Rotating Black Holes}
\author{Valeri P. Frolov and David Kubiz\v n\'ak}
\affiliation{Theoretical Physics Institute, University of Alberta,
Edmonton, Alberta, Canada, T6G 2G7}

\email{frolov@phys.ualberta.ca}

\email{kubiznak@phys.ualberta.ca}

\date{\today}

\begin{abstract}  
We demonstrate that the rotating black holes in an arbitrary
number of dimensions and without any restrictions on their rotation
parameters possess the same `hidden' symmetry as the $4$--dimensional
Kerr metric. Namely, besides the spacetime symmetries generated by
the Killing vectors they also admit the (antisymmetric) Killing--Yano
and symmetric Killing tensors. 
\end{abstract}

\pacs{04.70.Bw, 04.50.+h, 04.20.Jb \hfill 
Alberta-Thy-05-06}

\maketitle

The idea that the spacetime may have one or more large spatial
extra dimensions became very popular recently. In the brane world
models, which realize this idea, the usual matter is confined to the
brane, representing our world, while gravity propagates in the bulk.
Black holes, being the gravitational solitons, propagate in the bulk
and may be used as probes of extra dimensions. If the size $r_0$ of a
black hole is much smaller that the size $L$ of extra dimensions
and the black-hole--brane interaction is weak, the  black hole geometry is
distorted only slightly. This distortion is controlled by the
dimensionless parameter $r_0/L$. For many problems it is sufficient
to consider the limit when this parameter vanishes and approximate
the geometry by the metric of an isolated black hole. 
The metrics describing the isolated vacuum rotating higher dimensional black
holes were obtained by Myers and
Perry \cite{MP}. These solutions are the generalizations of the well
known $4$--dimensional Kerr geometry.
The symmetries play a key role in the study of physical
effects in the gravitational fields of black holes.
In this paper we demonstrate that the Myers--Perry
metrics besides the evident symmetries possess also an additional `hidden'
symmetry in the same way as it occurs for the Kerr spacetime. 

We start by reminding that the  Kerr metric possesses a number of
what was called by Chandrasekhar \cite{Chandra} `miraculous'
properties.  This metric was obtained by Kerr \cite{Kerr} as a
special solution which can be presented in the Kerr--Schild form
\be\n{KS} g_{\mu\nu}=\eta_{\mu\nu}+2H l_{\mu}l_{\nu}\, , \ee where
$\eta_{\mu\nu}$ is a flat metric and $l_{\mu}$ is a null vector, in
both metrics $g$ and $\eta$.   The Kerr solution is stationary and
axisymmetric, and it belongs to the metrics of the  special algebraic
type {\rm D}. Although the Killing vector fields $\pa_t$ and
$\pa_{\phi}$ are not enough to provide a sufficient number of
integrals of motion, Carter \cite{Carter} demonstrated that
both---the Hamilton--Jacobi and  scalar field equations---can be
separated in the Kerr metric. This `miracle' is directly connected
with the existence of an additional integral of motion associated
with the second rank Killing tensor \cite{WP} $K_{\mu\nu}=K_{(\mu\nu)}$ obeying the
equation 
\be\label{KT}
 K_{(\mu\nu;\lambda)}=0\, .
\ee 
As it was shown later, the equations for massless fields with
non--vanishing spin can be decoupled in this background, and the
variables separated in the resulting Teukolsky's master equations
\cite{Teuk_a,Teuk_b}.

Penrose and Floyd
\cite{Penrose}  demonstrated that the Killing tensor for the
Kerr metric can be written in the form $K_{\mu\nu}=f_{\mu\alpha}f^{\
\,\alpha}_{\nu}$, where the antisymmetric tensor
$f_{\mu\nu}=f_{[\mu\nu]}$ is the Killing--Yano (KY) tensor
\cite{Yano} obeying the equation $f_{\mu (\nu;\lambda)}=0$. Using
this object, Carter and McLenaghan  \cite{Carter4}, constructed the
symmetry operator of the massive Dirac equation. 

In many aspects a KY tensor is more fundamental than a Killing
tensor. Namely, its `square'   is always Killing tensor, but the
opposite is not generally true (see, e.g., \cite{Ferrando}).  In 
$4$--dimensional spacetime, as it was shown by Collinson \cite{Coll74}, 
if a vacuum solution of the Einstein equations allows a
non--degenerate KY tensor it is of the type D.  All the vacuum type D
solutions were obtained by Kinnersley \cite{Kinn}. Demianski and
Francaviglia \cite{DeFr} showed that in the absence  of the 
acceleration these solutions admit Killing and KY tensors.  It should
be  also mentioned that if a spacetime admits a non--degenerate KY
tensor it always has at least one Killing vector  \cite{Dietz}.

One can expect that at least some of these deep relations between
`hidden' symmetries and the algebraical structure of solutions of the
Einstein equations remain valid also in higher dimensional spacetimes.

Really, it was demonstrated that
the $5$--dimensional rotating black hole metric possesses the Killing
tensor and allows the separation of variables of the Hamilton--Jacobi
and scalar field equations \cite{FSa,FSb}. This separation
is also possible in higher  dimensional rotating black hole metrics
under a condition that their rotation parameters can be divided into
two classes, and within each of the classes the rotation  parameters
are equal one to another \cite{VSP}. Below we show that all the known
vacuum rotating black hole metrics in an arbitrary number of dimensions
and without any restrictions on their rotation parameters admit both
the Killing--Yano and the Killing tensors.

The Myers--Perry (MP) metrics \cite{MP} are the most general known
vacuum solutions for the higher dimensional rotating black holes
\cite{remark}. These metrics allow the Kerr--Schild form \eq{KS},
and, as it was shown recently \cite{Col}, they are of the type D.
The  MP solutions have slightly different form for the odd and even
number of spacetime dimensions $D$. We can write them compactly as
\begin{eqnarray}\n{MP}
ds^2&=&-dt^2+\frac{Udr^2}{V-2M}+\frac{2M}{U} (\,dt+
\sum_{i=1}^{n}a_i\mu_i^2d\phi_i)^{2} \non\\
&+&
\sum_{i=1}^{n}(r^2+a_i^2)(\mu_i^2d\phi_i^2
+d\mu_i^2)+\varepsilon r^2
d\mu^2_{n+\varepsilon}\, ,
\end{eqnarray} 
where
\be
V=r^{\varepsilon-2}\prod_{i=1}^{n}(r^2+a_i^2)\hhh
U=V(1-\sum_{i=1}^n\frac{a_i^2 \mu_i^2}{r^2+a_i^2})\, .
\ee
Here $n=[(D-1)/2]$, where $[A]$ means the integer part of $A$. We
define $\varepsilon$ to be $1$ for $D$ even and $0$ for odd.   The
coordinates $\mu_i$ are not independent. They obey the following
constraint
\begin{equation}\label{constraint}
\sum_{i=1}^n\mu_{\!i}^2+\varepsilon \mu^2_{n+\varepsilon}=1.
\end{equation} 
The MP metrics possess $n+1$ Killing vectors, $\pa_t$,
$\pa_{\phi_i}$, $i=1,\dots,n$. 

We show now that there is an additional 
symmetry connected with the KY tensors. 
The KY tensor is a special case of what is called a {\em conformal KY
tensor} which is defined as a $p$--form $k$ obeying the equation 
\cite{Tachibana, Kashiwada,Cari}
\ba
\nabla_{\!(\alpha_1}k_{\alpha_2)\dots \,\alpha_{p+1}}&=&
g_{\alpha_1\alpha_2} \Phi_{\alpha_3\dots \,\alpha_{p+1}}\non\\
&-&(p-1)\, g_{[\alpha_3 (\alpha_1} \Phi_{\alpha_2)\alpha_4\ldots
\alpha_{p+1}]}\, ,\\
\Phi_{\alpha_3\alpha_4\ldots \alpha_{p+1}}&=&
{1\over D+1-p} \nabla_{\!\beta} k^{\beta}_{\, \, \alpha_3\alpha_4\ldots
\alpha_{p+1}} \, .
\ea
KY tensors themselves form a subset of all conformal KY tensors for
which $\Phi=0$. Thus   the KY tensor of rank $p$ is a
$p$--form $f_{\alpha_1\dots \,\alpha_{p}}$ obeying the equation
\begin{equation}\label{Yeq}
\nabla_{\!(\alpha_1}f_{\alpha_2)\alpha_3\dots \,\alpha_{p+1}}=0\, .
\ee

In what follows we shall use the following properties. Denote by
$e_{\alpha_1\! \ldots \alpha_D}$ the totally  antisymmetric tensor
\be
e_{\alpha_1\!\ldots \alpha_D}=
\sqrt{-g}\, \epsilon_{\alpha_1\!\ldots \alpha_D}\, ,\, \,
e^{\alpha_1\!\ldots \alpha_D}=-{1\over
\sqrt{-g}}\, \epsilon^{\alpha_1\!\ldots \alpha_D}\, .
\ee
This tensor obeys the property ($q+p=D$)
\be\n{e_e}
e_{\mu_1\!\ldots \mu_q\beta_1\ldots\beta_p}
e^{\nu_1\!\ldots \nu_q\beta_1\ldots\beta_p}=-p!\, q!\, 
\delta_{\mu_1}^{[\nu_1}\ldots \delta_{\mu_q}^{\nu_q]}\, .
\ee
The Hodge dual $*\omega$ of the form $\omega$ is defined as
\be
(*\omega)_{\mu_1\,\ldots \mu_{D-p}}
={1\over p!}e_{\mu_1\,\ldots \mu_{D-p}}^{\quad \, 
\quad \quad\alpha_1\,\dots \,\alpha_p }\omega_{\alpha_1\,\dots \,\alpha_p
}\, .
\ee
One can check that $*(*\omega)=-\omega$. The Hodge dual of a
conformal KY tensor is again a conformal KY tensor. A conformal KY
tensor $k$ is dual to the KY tensor if and only if it is closed
$dk=0$  \cite{Cari}.

We focus our attention on the KY tensor $f$ of the rank $p=D-2$, so
that its Hodge dual $k=*f$ is the second rank conformal KY
tensor obeying the equations
\be\n{cye}
k_{\alpha\beta;\,\gamma}+k_{\gamma\beta;\,\alpha}
=\frac{2}{D-1} (g_{\alpha\gamma}\,k^{\sigma}_{\
\beta;\,\sigma}+ g_{\beta(\alpha}k_{\gamma)\, \, \,  ;\,\sigma}^{\,
\, \,  \,\sigma} )\,,
\ee
\be\n{dk}
k_{[\alpha\beta,\gamma]}=0\, .
\ee
The relation \eq{dk} implies that, at least locally, there
exists a one--form (potential) $b$ so that $k=db$.

We use the following ansatz for the conformal KY
potential $b$  for the MP metric \eq{MP}
\begin{equation}\label{cykp}
2b=(r^2+\sum_{i=1}^na_i^2\mu_i^2) \, dt+
\sum_{i=1}^na_i\mu_i^2(r^2+a_i^2)\, d\phi_i.
\end{equation}
The corresponding conformal KY tensor $k$ reads
\ba\n{cyk}
k&=&\sum_{i=1}^{n} a_i\mu_i\,d\mu_i\!\wedge\!
\left[a_idt+(r^2+a_i^2)\, d\phi_i\right] \non\\
&+&rdr\!\wedge\!  (dt+\sum_{i=1}^{n}a_i\mu_i^2d\phi_i )\, .
\ea
We emphasize that here and later on in similar formulas the
summation over $i$ is taken from $1$ to $n$ for both---even and odd number of
spacetime dimensions $D$; the coordinates $\mu_i$  are independent when $D$ is even 
whereas they obey the constraint \eq{constraint} when $D$ is odd.

To prove that $k_{\mu\nu}$ obeys \eq{cye} it is convenient to use the
Kerr--Schild form of the MP metrics. The required calculations are
straightforward but rather long. The details of the proof can be
found in \cite{FrKu}. For $D\le 8$ we also checked directly the
validity of the equation \eq{cye} by using the GRTensor program.

The Hodge dual of $k$,
\begin{equation}\n{fk}
f=-*k,
\end{equation} 
is the KY tensor. We shall give the explicit expressions for $f$ in
the dimensions 4 and 5. 

For $D=4$ (the Kerr geometry) there is
only one rotation parameter which, as usual, we denote by $a$. We also
put $\mu_1=\sin\theta$ and $\phi_1=\phi$. In these notations one recovers the standard
form of the Kerr metric and
\begin{eqnarray}
f^{(4)}&=&r\sin\theta d\theta\!\wedge\!\left[\,adt+(r^2+a^2)d\phi\right]\nonumber\\
&&-a\cos\theta dr\!\wedge\!\left(dt+a\sin^2\!\theta d\phi\right).
\end{eqnarray}
This expression coincides with the KY tensor discovered by Penrose
and Floyd \cite{Penrose}.

For $D=5$ there are 2 rotation parameters, $a_1$ and $a_2$. 
Using the constraint \eq{constraint} we write $\mu_2=\sqrt{1-\mu_1^2}$.
Thus we have
\begin{eqnarray}
-f^{(5)}&=&rdt\!\wedge\!dr\!\wedge
\left[a_2\mu_1^2d\phi_1+a_1(1-\mu_1^2)d\phi_2\right]\nonumber\\
&&+a_2\mu_1(r^2+a_1^2)dt\wedge d\mu_1\wedge d\phi_1\nonumber\\
&&-a_1\mu_1(r^2+a_2^2)dt\wedge d\mu_1\wedge d\phi_2\nonumber\\
&&+\mu_1(r^2+a_1^2)(r^2+a_2^2)\,d\phi_1\!\wedge\!d\phi_2\!\wedge\!d\mu_1\nonumber\\
&&+r\mu_1^2(1-\mu_1^2)(a_2^2-a_1^2)\,d\phi_1\!\wedge\!d\phi_2\!\wedge\!dr.
\end{eqnarray}
\begin{eqnarray}
-f^{(5)}&=&rdt\!\wedge\!dr\!\wedge
\left[a_2\mu_1^2d\phi_1+a_1(1-\mu_1^2)d\phi_2\right]\nonumber\\
&+&\mu_1dt\!\wedge d\mu_1\!\wedge
\left[a_2(r^2+a_1^2)d\phi_1-a_1(r^2+a_2^2)d\phi_2\right]\nonumber\\
&+&\mu_1(r^2+a_1^2)(r^2+a_2^2)\,d\phi_1\!\wedge\!d\phi_2\!\wedge\!d\mu_1\nonumber\\
&+&r\mu_1^2(1-\mu_1^2)(a_2^2-a_1^2)\,d\phi_1\!\wedge\!d\phi_2\!\wedge\!dr.
\end{eqnarray}

Using \eq{cyk} and \eq{fk} one can easily obtain $f$ in an explicit
form for an arbitrary number of dimensions. However, the rank of the
form $f$ grows with the number of dimensions and the corresponding 
expressions become quite long.

It is easy to check that the following object constructed from the KY
tensor,
\be\n{Kmn}
K_{\mu\nu}={1\over (D-3)!}\, f_{\mu\alpha_1\!\ldots\,\alpha_{D-3}}
f_{\nu}^{\, \, \alpha_1\!\ldots\,\alpha_{D-3}},
\ee
is the second rank Killing tensor.  Using (\ref{e_e}) one can express
$K_{\mu\nu}$ in terms of $k$
\be
K_{\mu\nu}=k_{\mu\alpha}k_{\nu}^{\, \,\alpha}
-{1\over 2}\, g_{\mu\nu}k_{\alpha\beta}k^{\alpha\beta}.
\ee
Evidently, $Q_{\mu\nu}=k_{\mu\alpha}k_{\nu}^{\, \,\alpha}$,
is the conformal Killing tensor, satisfying, $Q_{(\alpha\beta;\gamma)}=g_{(\alpha\beta}Q_{\gamma)}$,
where
\be
Q_{\gamma}=\frac{1}{D+2}\,(2Q^{\kappa}_{\ \gamma;\kappa}+Q^{\kappa}_{\ \kappa;\gamma}).
\ee
The calculations give the following expression for the Killing tensor
\begin{eqnarray}\n{kt}
K^{\mu\nu}&=&\sum_{i=1}^{n}\left[ a_i^2(\mu_i^2-1)\, g^{\mu\nu}
+a_i^2\mu_i^2\delta^{\mu}_t\delta^{\nu}_t+
\frac{1}{\mu_i^2}\,
\delta^{\mu}_{\phi_i}\delta^{\nu}_{\phi_i}\right] \nonumber\\
&+&\!\!\sum_{i=1}^{n-1+\varepsilon}\!\!\delta^{\mu}_{\mu_i}\delta^{\nu}_{\mu_i}
-2Z^{(\mu}Z^{\nu)}-2\xi^{(\mu}\zeta^{\nu)},
\end{eqnarray}
where 
\begin{equation}
\xi=\pa_t
\hhh \zeta=\sum_{i=1}^n a_i\pa_{\phi_i},\quad
Z=\sum_{i=1}^{n-1+\varepsilon}
\mu_i\partial_{\mu_i}. 
\end{equation}
The term constructed from the Killing
vectors $\xi^{(\mu}\zeta^{\nu)}$ can be excluded from  $K$.  In
the $4$--dimensional spacetime \eq{kt} reduces to the Killing tensor
obtained by  Carter \cite{Carter}, while in the $5$--dimensional case
it coincides with the Killing tensor obtained in \cite{FSa,FSb} after
the term  $\xi^{(\mu}\zeta^{\nu)}$ is omitted. 

The constructed KY and Killing tensors have direct connections with
the isometries of the background MP geometry. First of all, for a
second rank conformal KY tensor $k$ in a Ricci--flat spacetime 
\be
\xi^{\mu}={1\over D-1}\,k^{\sigma\mu}_{\ \ \ ;\,\sigma}
\ee
is a Killing vector \cite{Jez}. In particular, for the MP metrics
$\xi=\pa_t$. It is also easy to show that the vector
$\eta_{\mu}=K_{\mu\nu} \xi^{\nu}$  possesses the property
\be
\eta_{(\mu;\nu)}=-{1\over 2} {\cal L}_{\xi} K_{\mu\nu}\, .
\ee
For the MP metrics the Lie derivative of ${\cal L}_{\xi}K$ vanishes,
so that $\eta$ is a Killing vector. 
Calculations give
\be
\eta=\sum_{i=1}^{n}\!a_i^2\,\partial_t -\zeta\, .
\ee

The described `hidden' symmetry of a higher dimensional rotating
black hole implies the existence of an additional integral of motion.
For example, for a freely moving particle  with the velocity
$u^{\mu}$ the Killing tensor \eq{kt} implies that the quantity
$K_{\mu\nu}u^{\mu}u^{\nu}$ is constant. Because of the presence of
the KY tensor $f$ the classical spinning particles in the MP metric
possesses enhanced worldline supersymmetry \cite{Cari}. Similar
symmetries are also valid on the quantum level. In particular, the
operator $\nabla_{\!\mu}(K^{\mu\nu}\nabla_{\!\nu})$ commutes with the
scalar Laplacian $\Box=g^{\mu\nu}\nabla_{\!\mu}\nabla_{\!\nu}$
\cite{Carter3,Carter4}. Using the KY tensor it is possible to
construct an operator which commutes with the Dirac operator
\cite{Carter4,Benn,Cari}. In general  there exist deep relations
between KY tensors and the supersymmetry \cite{Gibb}.

It should be emphasized that for $D\ge 6$ the obtained  KY and
Killing tensors for the MP solutions do not guarantee the  separation
of variable in the Hamilton--Jacobi and Klein--Gordon equations.  This 
can be easily illustrated for a particle. The spacetime
symmetry of the $D=2n+1+\varepsilon$ dimensional MP metric
guarantees the existence of $n+1$ integrals of motion. The
normalization of the velocity gives one more integral. In order to
have the separability, there must exist at least
$J=D-(n+2)=n+\varepsilon-1$ additional integrals of motion. For $D=4$
and $D=5$ one has $J=1$, thus the Killing tensor which exists in the MP
metrics is sufficient for the separation of variables. For $D\ge 6$,
$J>1$, so that one cannot expect the separation of variables unless
there exist additional `hidden' symmetries. General conditions of the
separability of the Hamilton--Jacobi and Klein--Gordon equations were
obtained in \cite{DeFr}. Whether such additional symmetries exist in
the MP metrics is an interesting open question.

\noindent  

\section*{Acknowledgments}  
\noindent  
One of the authors (V.F.) thanks the Natural Sciences and Engineering
Research Council of Canada and the Killam Trust for the financial
support. The other author (D.K.) is grateful to the Golden Bell Jar
Graduate Scholarship in Physics at the University of Alberta.



\begin{thebibliography}{}
\bibitem{remark} In the higher dimensions there is no uniqueness
theorem and at the moment it is not known whether there exist
stationary vacuum black hole solutions, different from the
MP--metrics. Besides black holes with the horizon topology $S^{D-2}$ 
in the higher dimensions there exist other `black objects', like
black strings, and (at least for $D=5$) black rings \cite{BR}. These objects
may have properties quite different from the properties of the Kerr
metric. For example, the black ring metric is of type $I_i$ \cite{Col}.
\end{thebibliography}
\end{document}